# Shear Melting at the Crystal-Liquid Interface:

# Erosion and the Asymmetric Suppression of Interface Fluctuations

Malcolm Ramsay and Peter Harrowell

*School of Chemistry, University of Sydney, Sydney 2006 NSW, Australia*

Abstract

The influence of an applied shear on the planar crystal-melt interface is modelled by a nonlinear stochastic partial differential equation of the interface fluctuations. A feature of this theory is the asymmetric destruction of interface fluctuations due to advection of the crystal protrusions on the liquid side of the interface only. We show that this model is able to qualitatively reproduce the nonequilibrium coexistence line found in simulations. The impact of shear on spherical clusters is also addressed.

The relative stability of a crystal phase at its liquid melt is established, at equilibrium, by the relative magnitudes of the Boltzmann weighted configuration spaces for each phase. The problem of establishing the crystal-liquid coexistence point is thus well posed and, in the thermodynamic limit, independent of the mechanism of the transition or the characteristics of the interface between the two phases. This happy state vanishes, however, when we subject our system to some external gradient, such that one or both phases are maintained in a nonequilibrium steady state. In this paper we shall address the problem of coexistence of a crystal and its melt under the influence of an applied shear stress, one small enough that it induces only a linear response in either phase. This means that the crystal will be elastically strained while the liquid will experience a Newtonian shear flow. We consider the case of a planar interface between strained crystal and shearing liquid, one orientated such that the stress gradient is normal to the interface and the liquid flow is parallel to the surface. (The alternative, in which the stress gradient is parallel to the surface would result in a more complex (and physically ambiguous) situation with either no fluid flow or fluid flow characterised by two gradients.)

The combination of crystallization and shear flow crops up in a wide range of phenomena. Some of these, such as shear-induced crystallization of polymers [1], lyotropic solutions [2], colloidal suspensions [3] and metallic alloys [4] involve nonlinear effects associated with perturbation of the liquid structure. Nonlinear behaviour in the crystal phase includes yielding and fracture and, in the case of colloidal crystals, sliding crystals. Restricting ourselves to linear response in both liquid and crystal, we shall specifically address the suppression of crystallization by an applied shear, a phenomena that has been reported in colloids [5] and associated with wear between metal surfaces [6] and stick-slip lubrication in thin films [7]. Modelling of shear-induced melting of molecular crystals [8] has indicated that the phenomenon should be experimentally observable for viscous materials such as waxes (e.g. eicosane). We note that our restriction to linear response in the bulk phases does not preclude the possibility of nonlinear rheology of the two phase system due to shear-induced disordering [8].

A number of previous studies have addressed the problem of the nonequilibrium coexistence between strained crystal and flowing liquid. Ramaswamy and Renn [9] proposed that the depression of the freezing point was the result of the increase in the chemical potential of the crystal due to elastic strain. The fundamental assumption here was that a balance of chemical potentials described coexistence even when the liquid phase was flowing. In a simulation of a Lennard-Jones crystal and liquid under shear, Butler and Harrowell [10] found that the depression of the crystal-liquid coexistence point (as measured by the stability of a planar crystal-liquid interface) due to shear flow was substantially larger than could be accounted for by the elastic energy of the crystal proposed in ref. 9. The authors of ref. 10 found that no reasonable argument based on equating properties of the two bulk phases could account for the observed nonequilibrium coexistence line. Instead, it was proposed that it was the stability of the interface that was determining the coexistence point. The goal of this paper is to clarify this latter proposal and to develop a model capable of both articulating just how the shear flow influences the interface stability and to provide the basis for a quantitative treatment of this phenomenon.

We shall consider the case of 2D with a 1D interface. The generalization to 3D should be straight forward. Let the h(x) be the height of the interface at position x parallel to the interface and let the y > h(x) lie in the liquid phase and y < h(x) lie in the solid phase. The position of the crystal liquid interface has been shown to be accurately identified through the use of local order parameters such as that developed by Briels and Tepper for the face centred cubic crystal [11]. Karma [12] has demonstrated that, if temperature and density fluctuations are neglected, the fluctuations in the interface height at coexistence can be described by the following Langevin equation ,

$$\frac{1}{\xi}\frac{\partial h(x,t)}{\partial t} = \Delta + \Gamma h'' + \eta(x,t) \qquad (1)$$

where $\xi$ is the kinetic coefficient for interface motion, $\Delta = T_m - T$ (where $T_m$ is the equilibrium melting temperature) and $h'' = \partial^2 h/\partial x^2$. $\Gamma$ is the Gibbs-Thomson coefficient [13] which determines the degree to which the equilibrium melting point is altered by a non-zero curvature of the interface and is given by

$$\Gamma = \frac{T_M}{L_v}(\gamma + \partial^2\gamma/\partial\theta^2) \qquad (2)$$

where $L_v = L/V$ where L and V are the molar heat of fusion and volume of crystal, $\gamma$ is the surface tension and $\theta$ is the angle of the surface relative to an underlying crystal plane. Unlike the case of the liquid-vapour interface, the curvature of a crystal-liquid interface can change the interfacial free energy two ways: by increasing the surface area or by changing the orientation of the surface relative to the underling structure of the crystal. The noise $\eta$ is delta correlated in time and space, i.e.

$$<\eta(x_1,t_1)\eta(x_2,t_2)> = Q\delta(x_1-x_2)\delta(t_1-t_2) \qquad (3)$$

Note that each term on the right hand side of Eq. 1 has units of temperature so that the noise $\eta$ should be regarded as a local transient fluctuation in the temperature.

A physical value can be assigned to Q by the following argument. Consider the spatial Fourier transform $h_q(t)$ of the height $h(x,t)$. By an equipartition of energy argument [14], the equilibrium variance $<h_q^2>$ will satisfy

$$<h_q^2> = \frac{k_B T_m}{A(\gamma + \partial^2\gamma/\partial\theta^2)q^2} \qquad (4)$$

where A is an area per molecule. We can solve Eq. 1 for $<h_q^2>$ by taking space and time Fourier transforms to get

$$<h_q^2> = \frac{\xi Q}{2\Gamma q^2} \tag{5}$$

Equating Eqs. 4 and 5 and substituting Eq. 2 we have

$$Q = \frac{2k_B T_m^2}{A L_v \xi} \tag{6}$$

As we shall show, the noise amplitude Q acts as a simple scaling parameter of the interface height and supercooling. To make this clear we shall treat Q as an independent parameter of the model in the following analysis.

The essential characteristic of an applied shear stress σ is the asymmetry of its effect across the interface. On the solid side the result is a static elastic shear strain while in the liquid a Newtonian shear flow results, with a shear rate proportional to σ. We propose that the coupling between the interface and the shearing liquid is through the height fluctuations of the interface. Specifically, we argue that the protrusions of the interface into the liquid will yield at stresses considerably smaller than that exhibited by the bulk. Readers may be familiar with the general observation, referred to as the Hall-Petch effect [15], in which the yield stress is found to *increase* as the grain size decreases. This increase in strength is the result of the reduced influence of mobile dislocations with decreasing grain size. This trend, however, is reversed as the grain size continues to decrease. Below some crossover size (roughly 15nm or ~ 40 atoms in the case of copper [16]) the yield stress does indeed decrease [17], as we are proposing, with decreasing grain size. This is due to the increasing importance of shear-induced slip planes [17]. This nanometre scale is just the width expected to dominate the fluctuations of the crystal-liquid interface. Our physical picture of the coupling of the liquid shear flow to the crystal order is that the solid fluctuations that protrude into the liquid side are subject to mechanical disruption by the advection (i.e. slip planes) driven by the adjacent liquid, with the higher curvature fluctuations (i.e. smaller transverse dimension) exhibiting the greater susceptibility. As the effect of the applied stress must be independent of its sign, the lowest order contribution of the shear stress must be quadratic. Combining these elements, we propose the following expression for the interface fluctuation in the presence of an applied shear stress σ,

$$\frac{1}{\xi}\frac{\partial h(x,t)}{\partial t} = \Delta + \Gamma h'' + \kappa \sigma^2 H(h'')h'' + \eta(x,t) \qquad (7)$$

H(y) is the Heaviside step function [H(y) = 1 if y <0 and 0 if y ≥ 0], and it represents the essential nonlinearity of the model, a consequence of the asymmetry across the interface of the influence of the liquid shear flow. We shall neglect any changes in the parameters ξ and Γ due to the depression of the freezing point as these are not expected to be significant. Strictly, the supercooling Δ in Eq. 7 should be reduced by the applied stress due to the elastic deformation of the crystal so that $\Delta_\sigma = \Delta - \frac{\sigma^2 T_m}{2GL}$ where G is the crystal shear modulus. As we have previously [10] demonstrated that the magnitude of this contribution of the stress is small relative to the observed shift of the coexistence temperature due to shear we shall omit it from the following analysis and just use the unperturbed supercooling Δ with the equilibrium melting point $T_m$.

As the step function H(y) only permits contributions for $h'' < 0$, it follows that shear dependent term on the right hand side of Eq. 7 can only make a negative contribution to the crystal growth rate. Eq. 7, therefore, describes the erosion of the crystal interface protrusions into the liquid. The suppression of the interface fluctuations by the shear flow of the liquid acts like an additional contribution to the surface stiffness, but one acting on the negative curvature fluctuations only. We shall introduce the following reduced units: $z = x\Gamma^{-1/2}$ and $\tau = \xi t$, leaving us with the final equation,

$$\frac{\partial h(z,\tau)}{\partial \tau} = \Delta + h''[1 + \hat{\sigma}^2 H(h'')] + \eta(z,\tau) \qquad (8)$$

where $h'' = \frac{\partial^2 h}{\partial z^2}$ and, the rescaled shear stress $\hat{\sigma}$ is given by $\hat{\sigma}^2 = \frac{\kappa \sigma^2}{\Gamma}$. To calculate the average mobility of the interface we need to average over the extent of the interface (i.e. over $z$) space and over the noise. This double average is indicated by $<...>$. It follows from Eq. 5 that the steady state reduced growth rate $\upsilon$ is given by

$$\frac{\partial <h(z,\tau)>}{\partial \tau} = \upsilon = \Delta + \hat{\sigma}^2(\sigma) <h''>_{h''<0} \qquad (9)$$

where the average $<...>_{h''<0}$ refers to piecewise average along the interface taken only over those segments for which the curvature is negative (i.e. where the crystal protrudes into the liquid). Non-equilibrium coexistence corresponds to the locus of points in the space of $\hat{\sigma}$ and $\Delta$ for which $\upsilon = 0$, i.e.

$$\Delta_{coex} = -\hat{\sigma}^2_{coex} <h''>_{h''<0} \qquad (10)$$

We have numerically integrated Eq. 9 for a range of values of the supercooling $\Delta$ and the shear stress $\hat{\sigma}$. The spatial derivatives are treated using a second order central finite difference approximation with periodic boundary conditions using a spatial step length of $\delta x$, and the numerical integration of the time derivative is achieved using the Euler method with a step length $\delta t$. The noise is generated from a normal zero mean distribution with a standard deviation of one and then scaled by $Q_o^{1/2} = Q^{1/2} \left( \sqrt{\delta t / \delta x^2} \right)^{-1}$.

In Fig. 1 we plot the nonequilibrium coexistence curve $\Delta_{coex}/Q^{1/2}$ vs $\hat{\sigma}^2_{coex}$. For small $\hat{\sigma}$, the regime over which our model assumptions are based (i.e. neglect of elastic strain, non-Newtonian behaviour), we find the following empirical relation,

$$\Delta_{coex}/Q^{1/2} = \frac{0.4\hat{\sigma}^2}{1+0.228\hat{\sigma}^2} \qquad (11)$$

This curve represents the supercooling that is required to stabilize the solid-liquid interface in the presence of an applied shear stress. For parameter values above the curve we have crystal growth and below, we have melting arising as a result of the shear induced erosion of the interface. The fact that data for different values of Q fall on the same curve in Fig. 1 implies that the average magnitude of the interface height $h(z)$ scales as $Q^{1/2}$. That the nonequilibrium coexistence temperature is dependent on the amplitude of the noise is a consequence of the crucial role played by fluctuations in the shear melting mechanism described here. The data from nonequilibrium simulations of the Lennard Jones crystal and liquid under shear stress [10] have been included as an insert in Fig. 1. The simulation data exhibits the same relationship as predicted by the model presented here. This agreement provides support for the proposition, embodied in Eq. 7, that the nonequilibrium coexistence is, in the linear regime, determined by the stability of the interface subjected to the asymmetric destruction of interface fluctuations by the applied shear. A direct numerical comparison between simulation and theory must wait until an independent estimate of the parameter $\kappa$ in Eq.7 associated with the convective disruption of the interface fluctuations has been determined.

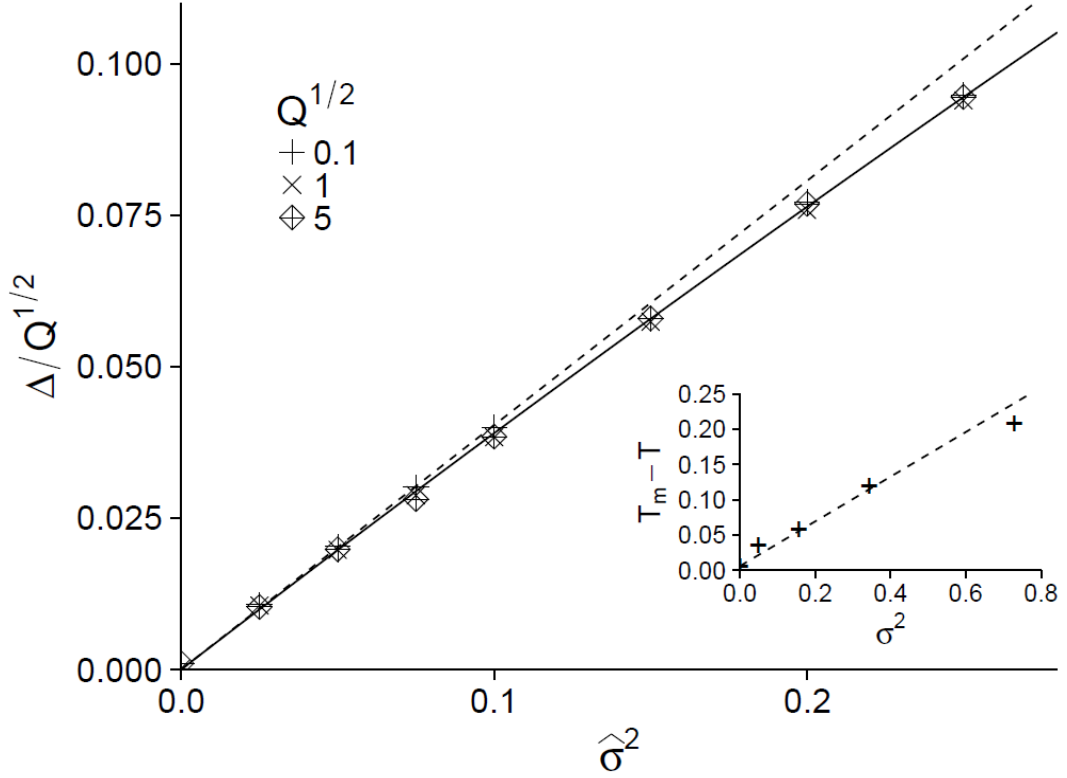

**Figure 1.** The nonequilibrium coexistence line in the space of the reduced supercooling $\Delta/Q^{1/2}$ and the square of the reduced shear stress $\hat{\sigma}^2$. The solid line corresponds to the empirical fit in Eq. 11 while the dashed line is a straight line fit to the data at small $\hat{\sigma}^2$, included for comaprison. Insert: The coexistence curve, plotted as $T_m$-T vs the squared shear stress $\sigma^2$ obtained from molecular dynamics simulations of the Lennard-Jones crystal in the presence of shear [10]. The dashed line is a straight line fit to the low stress data.

As is evident from Eq.10, the deviation from linearity between $\hat{\sigma}^2_{coex}$ and $\Delta_{coex}$ arises as a result of the influence of the applied shear stress on the interface fluctuations. Any suppression of interface fluctuations would reduce the coupling of the liquid shear to the interface and, hence, the degree to which the shear can disrupt crystallization. In Fig. 2, we plot the variance of the height fluctuations $<(h-<h>)^2>$ as a function of $\hat{\sigma}^2$ along the coexistence curve. Based on the implicit scaling of the height fluctuations with $Q^{1/2}$ concluded from the data in Fig. 1 we shall scale the height variance by Q. We find that the variance decreases roughly linearly with increasing $\hat{\sigma}^2$. Surprisingly, given the asymmetry of the action of the shear, we find very little asymmetry in the height fluctuations; the skewness $<(h-<h>)^3>/<(h-<h>)^2>^{3/2}$ remains less than 5% over this range of $\hat{\sigma}^2$. In the related problem of the influence of shear flow on the fluctuations of an interface between two fluids [18] with different viscosities, a similar reduction in interface fluctuations with increasing shear was reported.

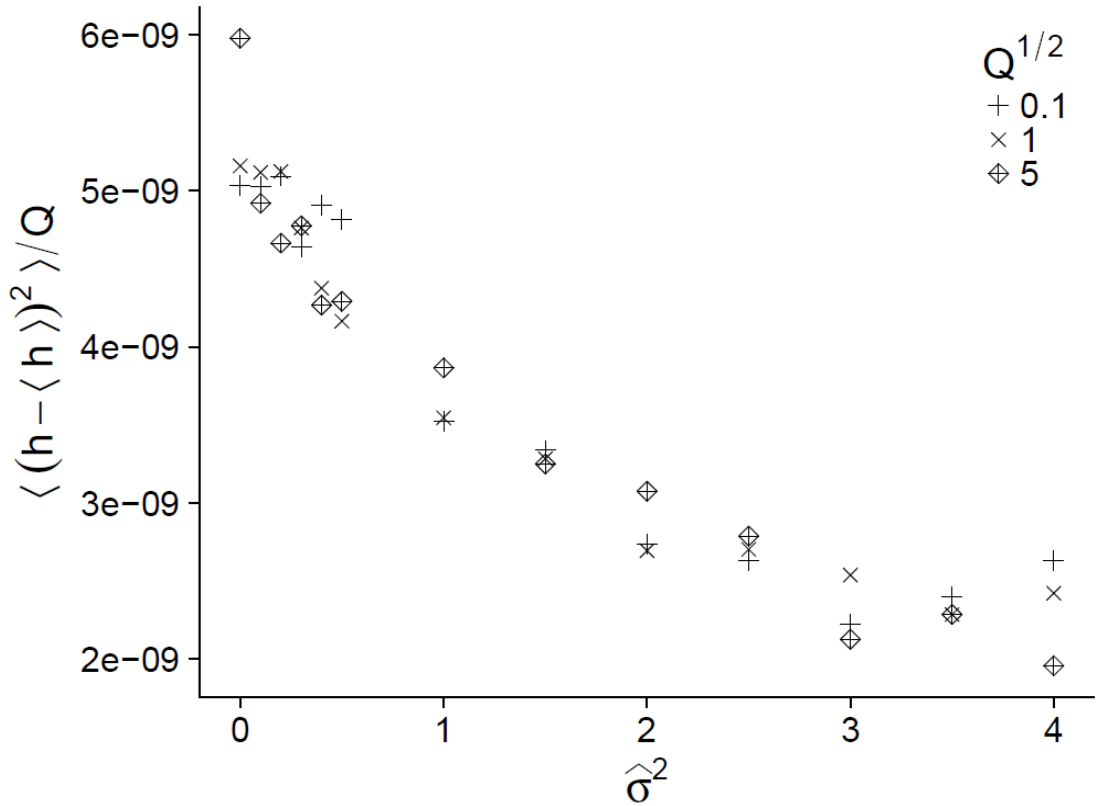

**Figure 2.** The variance $<(h-<h>)^2>/Q$ of the interface height fluctuations along the coexistence line as a function of $\hat{\sigma}^2$ for three different values of Q.

Away from coexistence, the interface velocity is given by Eq.9. If we approximate the actual value of $<h''>_{h''<0}$ by its value at coexistence for a given value of $\hat{\sigma}^2$, then we have

$$\upsilon = \Delta - \Delta_{coex}(\hat{\sigma}) \qquad (12)$$

where $\Delta_{coex}(\hat{\sigma})$ is the value of the supercooling at coexistence for the given stress $\hat{\sigma}$. In Fig. 3 we have plotted the $\upsilon$ against $\Delta - \Delta_{coex}(\hat{\sigma})$. We have scaled both quantities by $Q^{1/2}$ even though the scaling has no influence on the relation in Eq. 12 as a reminder that the absolute values of both the growth rate and the effective supercooling are dependent on the noise amplitude. We find that the approximate relation in Eq. 12 provides an excellent description of the dependence of the growth rate on supercooling and applied shear stress. This result suggests that the measurement of the steady state interface motion $\upsilon$ as a function of the supercooling $\Delta$ can, with Eq. 12, provide a useful strategy for evaluating $\Delta_{coex}(\sigma)$.

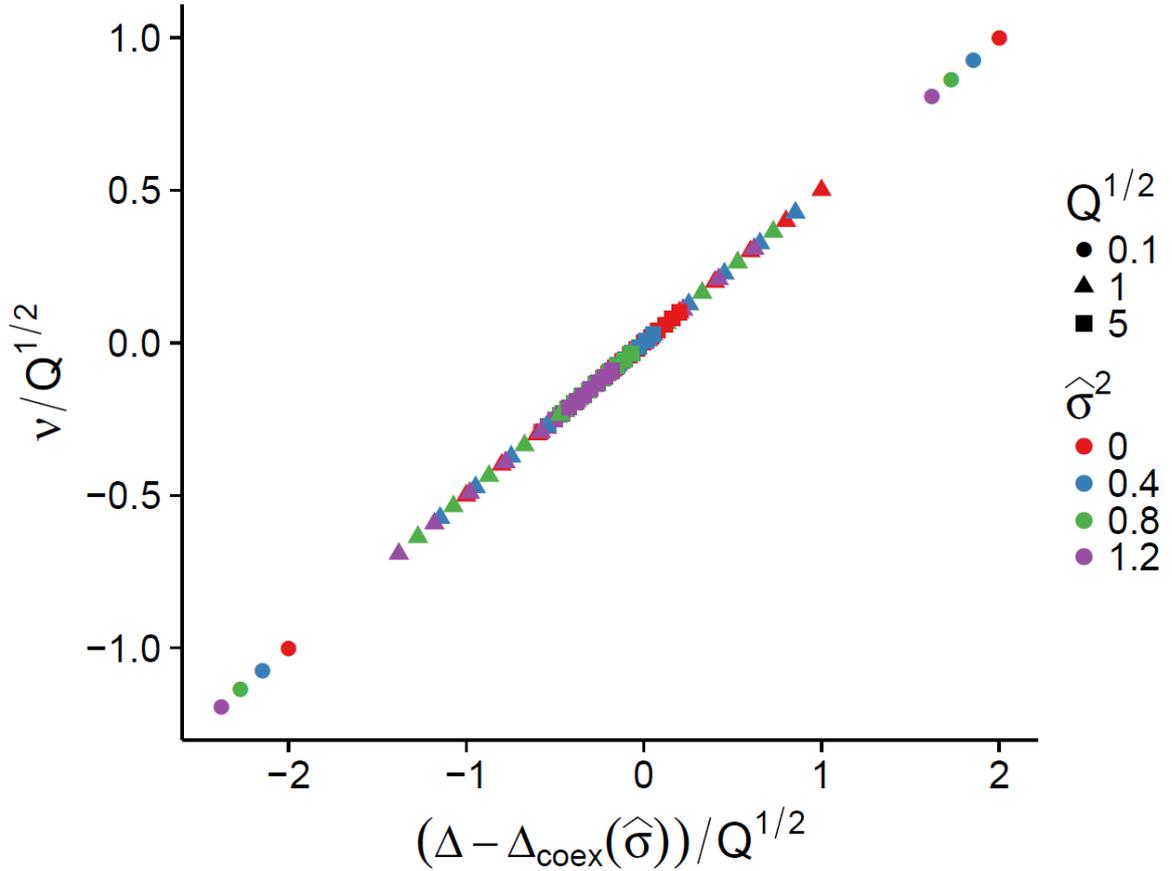

**Figure 3.** A plot of the crystal growth rate $\upsilon/Q^{1/2}$ against $(\Delta - \Delta_{coex}(\hat{\sigma}))/Q^{1/2}$ for a range of choices of the reduced stress $\hat{\sigma}$ and the noise amplitude $Q^{1/2}$.

In addition to the dissipation associated with shearing a liquid, any slip of the liquid against the solid surface represents an additional source of friction and, hence, dissipation. In the case of the nonequilibrium coexistence of crystal and shearing liquid described here, there must be an additional dissipation, that due to the work required to continually erode the crystal surface fluctuations. We propose that the rate of energy dissipation $\dot{\varepsilon}_{erosion}$ due to erosion along the nonequilibrium coexistence is given by

$$\dot{\varepsilon}_{erosion} = LA_s v_o(\hat{\sigma}) \approx -LA_s \Delta_{coex}(\hat{\sigma}) \qquad (13)$$

i.e. the latent heat of fusion L times the surface areas $A_s$ and the rate $v_o$ at which the shear stress $\hat{\sigma}$ disorders the crystal at $\Delta=0$. An interesting feature of this extra dissipation is that some of this energy can be recovered simply by reducing the shear stress to a point at which crystal growth can proceed with the associated release of the heat of fusion. In this sense, the shear disordering of the crystal represents an example of mechanical energy storage by a phase change material through the rectification provided by the asymmetry of the interface response to the applied shear.

How will erosion influence the stability of a spherical cluster of radius r? The shear stress experienced by a spherical particle in a planar shear flow of a Newtonian liquid has been considered by Bagster and Tomi [19]. The shear stress at the surface of a spherical cluster will vary with the angle of the surface normal with respect to the shear gradient in the gradient-velocity plane. The maximum shear stress corresponds to the point on the surface where the normal is parallel to the gradient. Here the shear stress is four times that of a planar surface $\sigma_p$ due to the distortion of the shear flow lines around the sphere. The spherical cluster will experience a torque that will accelerate it rotationally to an angular velocity of $\dot{\gamma}/2$. The fractional angular velocity $f$ as a function of time $t$ is

$$f = \frac{\Omega(t)}{\Omega_o} = 1 - \exp\left[-\frac{t}{\tau}\right] \text{ where } \tau = \frac{r^2 \rho}{15\eta} \text{ with } \rho, \text{ the mass density of the cluster.}$$

The maximum shear stress at the surface decreases as the cluster's angular velocity increases, according to the relation $\sigma_{max} = \sigma_p (4 - \frac{3f}{2})$, so that the steady state value of the maximum shear stress at the surface is $\sigma_{max} = \frac{5}{2}\sigma_p$. We note that experimental studies of the disruption of colloidal aggregates (flocs) by a shear flow have reported that disordering can occur either by erosion (a consequence of shear flow) and fragmentation (a consequence of the pressure difference) [20]

To summarise, spherical clusters will experience a larger shear stress and hence a greater rate of surface erosion in a shearing liquid than that of the planar surface under the same imposed shear gradient. While the magnitude of the maximum shear stress at the surface does not depend on the cluster size, the time required to reach the steady state angular velocity increases with the square of the cluster radius. This means that large clusters will experience a greater rate of erosion than smaller clusters. In addition to the enhanced shear stress at the surface of the rotating sphere over that of the planar interface, we note that the solidification rate $\Delta$, will also depend on the cluster radius r, decreasing with $r$ so as to vanish when it equals the radius of the critical nucleus. We conclude that when surface erosion precludes the possibility of a stable planar crystal interface at a given temperature and shear stress, it does likewise for all spherical clusters of the crystal.

In this paper we have presented an explicit model for the action of shear flow on the crystal-liquid coexistence based on the destruction of the crystal fluctuations into the liquid at the interface by advection. The model explains the observed large freezing point depression due to shear, as compared with the expectation based on the elastic contribution to the crystal chemical potential, by invoking the considerable reduction in the yield stress resulting from the small length scale of the interface fluctuations. We emphasise the significance of the asymmetry of the shear-induced erosion (expressed in Eq. 7 by the step function). If we were to replace the step function by a constant, i.e. allow the shear to destroy fluctuations equally on both the crystal and liquid sides of the interface, then the shear would simply constitute an augmented interfacial stiffness and would not produce any change to coexistence. In this sense the model provides a simple physical realization of the rectification of thermal fluctuations [21], one in which a system is driven by an asymmetric suppression of fluctuations. The shear induced disordering of the bulk crystal (i.e. without interface) involves a considerably larger shear stress than that considered here. This means that we expect a large hysteresis associated with shear melting a crystal, having disordered the crystal by a stress in excess of the yield value, the resulting liquid, on decreasing the stress, cannot crystallise until the stress has dropped below the critical value as determined by our interfacial mechanism. Direct experimental measurements of the nonequilibrium coexistence between a flowing liquid and the strained crystal are feasible. Previously we have presented a quantitative modelling of the parameters of this experiment for a variety of materials [8]. It is hoped that the theoretical treatment of crystal erosion presented here will stimulate interest in this fundamental phenomenon.

**Acknowledgements**

PH gratefully acknowledges valuable discussions with Thomas Voigtmann. This works has been supported by the Australian Research Council.